\documentclass [11pt]{article}

\usepackage{amsmath,amssymb,epsfig,cite,color,verbatim}
\usepackage{graphics,float}
\usepackage[colorlinks=true,citecolor=red,urlcolor=black,linkcolor=blue]{hyperref}
\usepackage{blindtext}
\usepackage{microtype}
\usepackage[latin9]{inputenc}
\usepackage{authblk}

\numberwithin{equation}{section}
\allowdisplaybreaks

\title{Born-Oppenheimer Quantization of the Matrix Model for $\mathcal{N}=1$ super-Yang-Mills}

\author[1]{Ver\'onica Errasti D\'iez\footnote{veroerdi@mppmu.mpg.de}}
\author[2]{Mahul Pandey\footnote{mpandey@stp.dias.ie}}
\author[3]{Sachindeo Vaidya\footnote{vaidya@cts.iisc.ernet.in}}

\affil[1]{ {\small {\it Max-Planck-Institut f\"ur Physik (Werner-Heisenberg-Institut),
F\"ohringer Ring 6, 80805 Munich, Germany}}}
\affil[2]{ {\small {\it School of Theoretical Physics, Dublin Institute for Advanced Studies,
10 Burlington Road, Dublin 4, Ireland}}}
\affil[3]{ {\small {\it Center for High Energy Physics, Indian Institute of Science,
Bangalore, 560012, India}}}

\date{}

\begin{document}
\maketitle

\begin{abstract}
We construct a quantum mechanical matrix model that approximates $\mathcal{N}=1$ super-Yang-Mills on $S^3\times\mathbb{R}$. 
We do so by pulling back the set of left-invariant connections of the gauge bundle onto the real superspace,
with the spatial $\mathbb{R}^3$ compactified to $S^3$.
We quantize the {$\mathcal{N}=1$ $SU(2)$ matrix model} in the weak-coupling limit using the Born-Oppenheimer approximation and find that different superselection sectors emerge for the effective gluon dynamics in this regime, reminiscent of different phases of the full quantum theory.
We demonstrate that the Born-Oppenheimer quantization is indeed compatible with supersymmetry, albeit in a subtle manner. In fact, we can define effective supercharges that relate the different sectors of the matrix model's Hilbert space.
{These effective supercharges} have a different definition in each phase of the theory.
\end{abstract}

\section{Introduction}
\label{intro}

It is hard to overemphasize the importance of Yang-Mills theory~\cite{YM} in theoretical physics.
Suffice to recall that it forms the basis of our understanding of the Standard Model of particle physics.
A reduction of Yang-Mills to a matrix model was obtained in~\cite{Balachandran:2014voa}.
By construction, this matrix model is free of divergences and captures relevant topological features of the full quantum field theory.
For instance
it has been successful in accounting for the impure nature of colored states in QCD~\cite{Balachandran:2014voa,Balachandran:2014iya}.
It has also been used to describe the low-lying glueball spectrum of pure QCD~\cite{Acharyya:2016fcn},
as well as to realize edge-localized glueball states in $SU(2)$ Yang-Mills theory~\cite{Acharyya:2017uhl}.

A remarkable property of this matrix model is that it exhibits a rich zero-temperature quantum phase structure when coupled to fermions.
This property was studied in detail in~\cite{Pandey:2016hat}, where it was observed in particular that the matrix model in~\cite{Balachandran:2014voa},
when weakly coupled to fundamental fermions and subsequently solved in the Born-Oppenheimer approximation,
demonstrates quantum critical behaviour at special corners of the gauge configuration space.
It was also shown that this model is suitable to describe the Yang-Mills regime with zero temperature and large baryon chemical potential.
In this regime, the coupling is weak and the quarks can be thought of as forming a Fermi sea,
with only a few energy levels available near the Fermi surface~\cite{Alford}.
Then the phases of quark matter are typically characterized by spatially uniform fermion condensates.
Consequently, a quantum mechanical matrix model (as the one described in~\cite{Balachandran:2014voa}) is an appropriate starting point to study the key features of such condensates. Indeed, the phases of the matrix model were found to show a similar symmetry-breaking pattern as the color-spin-locked phase
conjectured in one-flavor QCD~\cite{Schafer}.

Supersymmetric extensions of Yang-Mills theories have received considerable theoretical attention too, starting with~\cite{Ferrara:1974pu}.
For $\mathcal{N}<4$, a noteworthy aspect of these supersymmetric theories resides in their non-trivial vacuum structure.
In this context, the groundbreaking finding of the exact low energy effective action and spectrum of BPS states for $\mathcal{N}=2$,
$SU(2)$ super-Yang-Mills stands out~\cite{Seiberg:1994rs}.
On the other hand, the investigation of the vacuum structure and domain wall configurations of the $\mathcal{N}=1$,
$SU(N)$ analogous theories continues to be pursued with zeal, see e.g.~\cite{Kovner:1997ca}.

In this paper, we focus on the $\mathcal{N}=1$ supersymmetric extension of the $SU(2)$ matrix model
and study its quantum phase structure in the weak-coupling regime via Born-Oppenheimer quantization.
We find two distinct quantum phases for the effective gauge dynamics.
Our construction can be easily generalized to arbitrary $SU(N)$
and allows for the inductive inference of matrix models with extended supersymmetry. 
For concreteness, the $\mathcal{N}=2$ counterpart to our matrix model follows from coupling
the $\mathcal{N}=1$ gauge multiplet $(M_{\mu a},\lambda_\alpha^a,D^a)$ introduced in (\ref{WinWZ})
to an $\mathcal{N}=1$ matter multiplet and then imposing an internal $SU(2)$ symmetry.

At first sight, it may seem that the Born-Oppenheimer quantization is not quite justified in this case,
as it treats the gauge fields and fermions on unequal footing.
Thus, such a quantization procedure a priori seems inconsistent with supersymmetry.
Indeed, there is no apparent supersymmetry in the effective Hamiltonian governing the gauge field dynamics.
However, by a careful analysis, we demonstrate that supersymmetry gets restored in the full Hilbert space,
which is the direct sum of the Hilbert spaces of all possible effective Hamiltonians
corresponding to fermions occupying different energy levels.

The paper is organized as follows.
In section \ref{sec2},
we derive a quantum mechanical matrix model for $\mathcal{N}=1$ supersymmetric Yang-Mills theory with gauge group $SU(2)$.
We do so by extending the matrix model~\cite{Balachandran:2014voa} to the real $\mathcal{N}=1$ superspace.
In more detail, we begin by briefly reviewing the construction of the non-supersymmetric matrix model in section \ref{secrev}.
We then introduce the relevant superspace in section \ref{secspace}, thereby setting our notation.
After obtaining the corresponding superconnection in section \ref{secconnection},
we adapt Sohnius' maximal approach~\cite{Sohnius} to suitably constrain this superconnection
and infer the action of the supersymmetric matrix model in section \ref{secaction}.
Section \ref{sechamiltonian} is devoted to the Hamiltonian formulation of the matrix model, including the algebra of the supercharges.
Next, section \ref{secBornOpp} describes the Born-Oppenheimer quantization of the supersymmetric matrix model.
This is done in two steps.
First, we find the fermionic spectrum in section \ref{sec31}.
Second, we compute the effective gauge dynamics induced by the fermions near the Fermi surface in section \ref{sec32}.
We thus demonstrate that there are two distinct phases for the gluons.
In section \ref{sec4}, we describe how supersymmetry can be reconciled with the Born-Oppenheimer approximation,
by defining operators called effective supercharges which connect different sectors of the full Hilbert space.
We conclude in section \ref{sec5}, with a summary and discussion of our results.

\section{The supersymmetric matrix model}
\label{sec2}

In this section, we provide the $\mathcal{N}=1$ supersymmetric extension of the quantum mechanical matrix model for $SU(N)$ Yang-Mills theory introduced in~\cite{Balachandran:2014voa}.
This matrix model is in turn based on the canonical study of the said field theory with gauge group $SU(2)$ and defined on $S^3\times\mathbb{R}$
that was carried out in~\cite{NarRam}.
Although our construction is straightforwardly generalizable to arbitrary $N$,
for concreteness we will explicitly provide the supersymmetric matrix model for $N=2$ only.

\subsection{The matrix model: a review}
\label{secrev}

The main ideas involved in the definition of the non-supersymmetric matrix model with gauge group $SU(N)$ are as follows.
Start with the Maurer-Cartan left-invariant one form on $SU(N)$,
\begin{equation}
\Omega = {\rm Tr} \left( T_a u^{-1} d u\right)M_{ab} T_b, \qquad u \in SU(N).
\end{equation}
Here, ${T_a}$ are the generators in the defining representation of $\mathfrak{su}(N)$ and $M_{ab}$ is a real square matrix of order $N^2-1$.
Throughout the paper, we employ the normalization convention ${\rm Tr}(T_aT_b )=\delta_{ab}$. Next, consider isomorphically mapping the spatial $S^3$ onto an $SU(2)$ subgroup of the $SU(N)$ gauge group.
$X_i$, the three generators of translations on $S^3$,  are identified with the corresponding subset of generators $T_i$, with $i=1,2,3$.
The connection on $S^3$ is obtained by pulling back $\Omega$ under this map:
\begin{align}
-iA_i\equiv \Omega(X_i)=-iM_{ia}T_a, \label{Anonsusy}
\end{align}
where $M_{ia}$ is a rectangular $3\times(N^2-1)$-dimensional matrix that depends solely on time.
In other words, $A_i$ plays the role of the vector potential in the matrix model reduction of $SU(N)$ Yang-Mills theory.
Notice that we choose to work with a Hermitian connection: $A_i^\dagger=A_i$.
The homologous pullback of the structure equation for $\Omega$ yields the curvature $F_{ij}$ on $S^3$:
\begin{align}
F_{ij}\equiv d\Omega(X_i,X_j)+\Omega(X_i)\wedge\Omega(X_j). \label{Fnonsusy}
\end{align}

The supersymmetric matrix model we propose is obtained by generalizing the above procedure to $\mathcal{N}=1$ superspace.
We begin our construction by convening the basics of this superspace,
together with the relevant functions and differential operators defined on it.
This helps us establish our notation and conventions.

\subsection{The $\mathcal{N}=1$ superspace}
\label{secspace}

As is well-known, the $\mathcal{N}=1$ super-Poincar\'e algebra can be naturally realized on the super-Minkowski space $\mathbf{M}^{4|4}$,
which is identified with the coset space Super-Poincar\'e/Lorentz~\cite{Sohnius}.
A convenient parametrization of this coset space is given by the real (or symmetric) superspace, with coordinates
\begin{equation}
z^A\equiv\{x^\mu,\theta^\alpha,\overline{\theta}_{\dot{\alpha}}\}, \label{za}
\end{equation}
where the four-vector index $\mu$ ranges from zero to three and the spinor indices $(\alpha,\dot{\alpha})$ run from one to two.
Note that, while $(A,\mu,\alpha)$ are upper indices, $\dot{\alpha}$ is a lower index.
The $x^\mu$ are the usual bosonic coordinates on Minkowski spacetime, where we use the mostly positive metric $\eta_{\mu\nu}=diag(-1,1,1,1)$.
On the other hand, $(\theta^\alpha, \overline{\theta}_{\dot{\alpha}})$ span the fermionic directions of the superspace
and so they are Grassmann numbers.
For these anticommuting numbers, all usual spinor identities hold.
In particular, $(\theta^\alpha)^2=0=(\overline{\theta}_{\dot{\alpha}})^2$.
Spinor indices are raised/lowered with $2\times 2$ totally antisymmetric tensors, normalized so that
\begin{equation}
\quad\quad \epsilon^{\alpha\beta}=-\epsilon^{\dot{\alpha}\dot{\beta}}=i\sigma^2,
\end{equation}
with $\sigma^2$ the second Pauli matrix.
Subsequently, we compactify the bosonic spatial $\mathbf{R}^3$ to $S^3$.

Superfields are functions of superspace: $\phi=\phi(x^\mu,\theta^{\alpha},\overline{\theta}_{\dot{\alpha}})$.
As such, they are most conveniently expressed in terms of their (finite) power series expansion
in the Grassmannian directions $(\theta^\alpha, \overline{\theta}_{\dot{\alpha}})$.
If we indicate the superfields that generate translations on the superspace by $X_A\equiv \{X_\mu,X_\alpha,\overline{X}^{\dot{\alpha}}\}$,
then it follows that they satisfy the super-Poincar\'{e} algebra
\begin{align}
\nonumber
[X_i,X_j]&=\epsilon_{ijk}X_k, &
[X_i,X_\alpha]&=-\frac{1}{2}(\sigma_i)_{\alpha\dot{\alpha}}(\overline{\sigma}^0)^{\dot{\alpha}\beta}X_\beta, \\ 
[X_i,\overline{X}^{\dot{\alpha}}]&=\frac{1}{2}\overline{X}_{\dot{\beta}}(\overline{\sigma}^0)^{\dot{\beta}\alpha}
(\sigma_i)_{\alpha\dot{\gamma}}\epsilon^{\dot{\gamma}\dot{\alpha}}, &
\{X_\alpha,\overline{X}^{\dot{\alpha}}\}&=(\sigma^\mu)_{\alpha\dot{\beta}}\epsilon^{\dot{\beta}\dot{\alpha}}X_\mu,
\end{align}
where the sigma matrices form a basis for $2\times 2$ complex matrices and are defined as
\begin{eqnarray}
\sigma^\mu\equiv(-\mathbf{1},\sigma^i), \qquad
\overline{\sigma}^\mu=(-\mathbf{1},-\sigma^i), \label{sigmadef}
\end{eqnarray}
with $\mathbf{1}$ the identity matrix and $\sigma^i$ the Pauli matrices.
Notice these are related by a simple operation of index raising:
$(\overline{\sigma}^\mu)^{\dot{\alpha}\alpha}=\epsilon^{\dot{\alpha}\dot{\beta}}\epsilon^{\alpha\beta}(\sigma^\mu)_{\beta\dot{\beta}}$.

Additionally, we introduce the linear differential operators $D_A\equiv (\partial_\mu,D_\alpha,\overline{D}^{\dot{\alpha}})$ on the superspace, where 
\begin{align}
D_\alpha=\frac{\partial}{\partial \theta^\alpha}
-i(\sigma^\mu)_{\alpha\dot{\alpha}}\overline{\theta}^{\dot{\alpha}}\partial_\mu, \qquad
\overline{D}^{\dot{\alpha}}=\frac{\partial}{\partial\overline{\theta}_{\dot{\alpha}}}
+i\theta^\alpha(\sigma^\mu)_{\alpha\dot{\beta}}\epsilon^{\dot{\beta}\dot{\alpha}}\partial_\mu. \label{Dsdef}
\end{align}
Here, the differentials along the Grassmannian directions are defined by
\begin{align}
\frac{\partial}{\partial \theta^\alpha}\theta^\beta=\delta^\beta{}_\alpha, \qquad
\frac{\partial}{\partial\overline{\theta}_{\dot{\alpha}}}\overline{\theta}_{\dot{\beta}}=-\delta_{\dot{\beta}}{}^{\dot{\alpha}}.
\end{align}
The only non-vanishing (anti-)commutation relation between the $D_A$'s is given by
\begin{equation}
\{D_\alpha,\overline{D}^{\dot{\alpha}}\}=(\sigma^\mu)_{\alpha\dot{\beta}}\epsilon^{\dot{\beta}\dot{\alpha}}\partial_\mu. \label{ant}
\end{equation}
Observe that the differential operators above are defined so that they (anti-)commute with supertranslations: $[D_A,X_B\}=0$,
where the so-called graded commutator $[f,g\}$ denotes either commutator or anticommutator, according to the even or odd character of $f$ and $g$.
Thus the $D_A$'s should be regarded as the covariant derivatives
on the superspace\footnote{The covariant derivatives $D_A$ should not be confused
with the gauge-covariant derivatives $\nabla_A$ introduced in section \ref{secaction}.}.
Further, for any Lagrangian to be $\mathcal{N}=1$ supersymmetric, its kinetic term must be a function of the $D_A$'s.

There are three important classes of superfields that will be crucial in section \ref{secaction}.
Vector superfields $V$ satisfy the reality condition $V=V^\dagger$.
Chiral $\Phi$ and anti-chiral $\overline{\Phi}$ superfields are characterized by $\overline{D}^{\dot{\alpha}}\Phi=0$ and $D_\alpha\overline{\Phi}=0$, respectively.

\subsection{The superconnection}
\label{secconnection}

We are now ready to obtain the superconnection $\mathcal{A}_A$ on the above introduced $\mathcal{N}=1$ superspace.
We will do so in direct analogy to (\ref{Anonsusy}) before, while restricting attention to the gauge group $SU(2)$.
Namely, we shall isomorphically map this gauge group to the real superspace
and after that pullback its left-invariant one-form under the said isomorphism.

Consider a set of three real superfields $\phi_a(x^\mu,\theta^{\alpha},\overline{\theta}_{\dot{\alpha}})$, with $a=1,2,3$.
These enable us to define the following map from the real superspace to $SU(2)$:
\begin{equation}
\mathbf{M}^{4|4}\ni z^A\equiv (x^\mu,\theta^\alpha,\overline{\theta}_{\dot{\alpha}})\mapsto
g=\exp[i\phi_a(x^\mu,\theta^\alpha,\overline{\theta}_{\dot{\alpha}})T_a] \in SU(2),
\label{map}
\end{equation}
with $T_a$ the generators of the gauge group.
As usual, multiplication of group elements $g$ induces a motion in the $z^A$ parameter space: these are precisely the $X_A$ supertranslations.
The Maurer-Cartan form $\Omega=g^{-1}dg$ can then be pulled back under the map (\ref{map}) onto the $\mathcal{N}=1$ superspace,
thereby yielding the desired superconnection:
\begin{equation}
-i\mathcal{A}_A\equiv\Omega(X_A)=-i\mathcal{M}_{Aa}T_a. \label{superconn}
\end{equation}
The above real rectangular matrices $\mathcal{M}_{Aa}$ are functions of only time and the fermionic coordinates:
$\mathcal{M}_{Aa}=\mathcal{M}_{Aa}(t,\theta^\alpha,\overline{\theta}_{\dot{\alpha}})$.

Paralleling the derivation of the non-supersymmetric curvature in (\ref{Fnonsusy}),
we pullback the structure equation associated to $\Omega$ to obtain the curvature on $\mathbf{M}^{4|4}$:
\begin{align}
\mathcal{F}_{AB} \equiv d\Omega(X_A,X_B)-i[\Omega(X_A),\Omega(X_B)\}
=X_A(\mathcal{A}_B)-(-1)^{AB}X_B(\mathcal{A}_A)-i\Omega([X_A,X_B\})-i[\mathcal{A}_A,\mathcal{A}_B\}.
\end{align}
Explicitly, the various components of this curvature on the real superspace are given by
\begin{align}
\nonumber
\mathcal{F}_{0i}&=\partial_0\mathcal{A}_i-i[\mathcal{A}_0,\mathcal{A}_i], &
\mathcal{F}_{ij}&=-\epsilon_{ijk}\mathcal{A}_k-i[\mathcal{A}_i,\mathcal{A}_j],  \\
\nonumber
\mathcal{F}_{0\alpha}&=\partial_0\mathcal{A}_\alpha-D_\alpha \mathcal{A}_0-i[\mathcal{A}_0,\mathcal{A}_\alpha], &
\mathcal{F}_{i\alpha}&=-D_\alpha \mathcal{A}_i
+\frac{1}{2}(\sigma_i)_{\alpha\dot{\alpha}}(\overline{\sigma}^0)^{\dot{\alpha}\beta}
\mathcal{A}_\beta-i[\mathcal{A}_i,\mathcal{A}_\alpha],  \\
\nonumber
\mathcal{F}_{0\dot{\alpha}}&=\partial_0\mathcal{A}_{\dot{\alpha}}-
\overline{D}_{\dot{\alpha}}\mathcal{A}_0-i[\mathcal{A}_0,\mathcal{A}_{\dot{\alpha}}], &
\mathcal{F}_{i\dot{\alpha}}&=-\overline{D}_{\dot{\alpha}}\mathcal{A}_i
-\frac{1}{2}\mathcal{A}_{\dot{\beta}}(\overline{\sigma}^0)^{\dot{\beta}\alpha}(\sigma_i)_{\alpha
\dot{\alpha}}-i[\mathcal{A}_i,\mathcal{A}_{\dot{\alpha}}],  \\
\nonumber
\mathcal{F}_{\alpha\beta}&= D_\alpha \mathcal{A}_\beta+D_\beta \mathcal{A}_\alpha
-i\{\mathcal{A}_\alpha,\mathcal{A}_\beta\}, &
\mathcal{F}_{\dot{\alpha}\dot{\beta}}&=\overline{D}_{\dot{\alpha}}\mathcal{A}_{\dot{\beta}}+
\overline{D}_{\dot{\beta}}\mathcal{A}_{\dot{\alpha}}-i
\{\mathcal{A}_{\dot{\alpha}},\mathcal{A}_{\dot{\beta}}\}, \\
\mathcal{F}_{\alpha\dot{\beta}}&= D_\alpha \mathcal{A}_{\dot{\beta}}+\overline{D}_{\dot{\beta}}\mathcal{A}
_\alpha-(\sigma^\mu)_{\alpha\dot{\beta}}
\mathcal{A}_\mu-i
\{\mathcal{A}_\alpha,\mathcal{A}_{\dot{\beta}}\}, & {}&{}
\end{align}
with $(D_\alpha,\overline{D}_{\dot{\alpha}})$ as defined in (\ref{Dsdef}).

\subsection{The action}
\label{secaction}

Having established what the covariant derivatives, the superconnection and its curvature are in the $\mathcal{N}=1$ superspace,
we now proceed to construct the supersymmetric quantum mechanical matrix model of interest from these.
Recall that, for gauge theories in flat space, a gauge invariant Lagrangian is obtained by direct gauge-covariantization.
Namely, by replacing all spatial derivative operators $\partial_\mu$ in the Lagrangian by gauge-covariant derivative operators
$\nabla_\mu\equiv\partial_\mu-i A_\mu$, with $A_\mu$ the gauge field.
For any gauge group $G$ with generators $T_a$, a generic element is expressed as $u=\exp(i\varphi_a T_a)$,
where $\varphi_a$ are real parameters that depend on the flat space coordinates: $\varphi_a=\varphi_a(x^\mu)$.
In our conventions, gauge transformations act on gauge fields via $A_\mu\rightarrow uA_\mu u^{-1}$.
Equivalently, one says that gauge fields transform under the adjoint representation of the gauge group.
Meanwhile, the Lagrangian remains invariant.
As a consequence, the gauge group action does not fully specify gauge fields.
This freedom can be used to set constraints (or gauge-fixing conditions)
on the gauge fields and thus remove redundancies in the description of the theory.

The generalization of the just described approach to supersymmetric gauge theories is morally straightforward.
For each of the superspace covariant derivatives $D_A$, one introduces a gauge superpotential $\mathcal{A}_A$
and forms a gauge-covariant derivative $\nabla_A\equiv D_A-i \mathcal{A}_A$.
In general, it is not possible to make the a priori assumption that the $\mathcal{A}_A$'s are real;
they must be regarded as arbitrary complex superfields.
Accordingly, the gauge group elements $g=\exp(i\phi_aT_a)$ are to be taken as complex superfields themselves,
with $\phi_a=\phi_a(x^\mu,\theta^\alpha,\overline{\theta}_{\dot{\alpha}})$.
Again, the gauge group action on the gauge superpotentials leads to more degrees of freedom than required to describe the supersymmetric theory,
the gauge parameter superfields $\phi_a$ having too few components to be able to gauge away all these redundancies.
It follows that one must impose a set of gauge- and super-covariant constraints on the complex $\mathcal{A}_A$'s
to get rid of the irrelevant degrees of freedom.
However, which constraints to impose is a non-trivial decision
that was first elucidated in a geometrically consistent manner in~\cite{Gates:1983nr}.
This work paved the way to the now standard constraint procedure for obtaining a supersymmetric Yang-Mills theory from the superfields,
known as the maximal approach and thoroughly explained by Sohnius~\cite{Sohnius}.
In the following, we adapt the maximal approach to our setup.

Indeed, our matrix model reduction on $\mathcal{N}=1$ superspace has led us to exactly the same situation:
for each superspace covariant derivative $D_A$ (\ref{Dsdef}),
we have a corresponding gauge superpotential $\mathcal{A}_A= \mathcal{M}_{A a}T_a$ (\ref{superconn}).
Combining these, we define the gauge-covariant derivatives in the matrix model as
\begin{align}
\nabla_A\equiv D_A-i\mathcal{M}_{A a}T_a. \label{gaugecovder}
\end{align}
The intrinsically complex superfields $\mathcal{M}_{A a}=\mathcal{M}_{A a}(t,\theta^\alpha,\overline{\theta}_{\dot{\alpha}})$
have more components than needed to specify the super-Yang-Mills matrix model, so we must impose a consistent set of constraints on them.

First, we impose the constraints
\begin{eqnarray}
\mathcal{F}_{\alpha\dot{\alpha}}=0, \qquad
\mathcal{F}_{\alpha\beta}=0=\mathcal{F}_{\dot{\alpha}\dot{\beta}}.
\label{reppcstr}
\end{eqnarray}
The leftmost equation results from a simple field redefinition of the gauge superpotential.
The rightmost equations can be justified by arguments of consistency.
Supersymmetry requires coupling the gauge theory to matter.
In particular, consider couplings to chiral and anti-chiral superfields.
For the (anti-)chirality conditions ---displayed at the end of section \ref{secspace}--- to be compatible with gauge symmetry,
the rightmost equalities must be satisfied.
Notice that these constraints imply that both $\mathcal{M}_{\alpha a}T_a$ and $\mathcal{M}_{\dot{\alpha} a}T_a$ are pure gauge.
We make use of this gauge freedom to set
\begin{align}
\mathcal{M}_{\dot{\alpha} a}=0.
\end{align}
It is easy to check that, for our above (partial) gauge choice, 
\begin{equation}
\mathcal{M}_{\mu a}=-\frac{1}{2}(\overline{\sigma}_\mu)^{\dot{\alpha}\alpha}\overline{D}_{\dot{\alpha}}\mathcal{M}_{\alpha a} \label{Mmua}
\end{equation}
solves all constraints in (\ref{reppcstr}).

Further constraints are still needed.
Expressly, one must ensure the uniqueness of (\ref{Mmua}); i.e.~its real and imaginary parts should not be independent.
To this aim, note that the non-zero curvatures on the real superspace
fulfill the Bianchi identities $\nabla_{[\mu}\mathcal{F}_{\nu A\}}=0$ by construction.
It follows~\cite{BianchiSohnius} that all these curvatures can be expressed in terms of two superfields $W$ and $\overline{W}$ as
\begin{equation}
\mathcal{F}^a_{\mu\alpha}=(\sigma_\mu)_{\alpha\dot{\alpha}}\overline{W}_{\dot{\beta}a}\epsilon^{ \dot{\beta}\dot{\alpha}}, \quad
\mathcal{F}^a_{\mu\dot{\alpha}}=\epsilon^{\alpha\beta}W_{\beta a} (\sigma_\mu)_{\alpha\dot{\alpha}}, \quad
\mathcal{F}_{\mu\nu}^a=-\frac{1}{2}\left(\nabla_\alpha(\sigma_{\mu\nu})_\beta{}^{\alpha}\epsilon^{\beta\gamma}W_{\gamma a}
+\overline{\nabla}_{\dot{\alpha}}(\overline{\sigma}_{\mu\nu})^{\dot{\alpha}}{}_{\dot{\beta}}\overline{W}_{\dot{\gamma}a}\epsilon^{\dot{\gamma}\dot{\beta}}\right),
\end{equation}
where, based on the sigma matrices in (\ref{sigmadef}), we have defined
\begin{eqnarray}
\sigma_{\mu\nu}\equiv\frac{1}{2}(\sigma_\mu\overline{\sigma}_\nu-
\sigma_\nu\overline{\sigma}_\mu), \qquad \overline{\sigma}_{\mu\nu}\equiv\frac{1}{2}
(\overline{\sigma}_\mu\sigma_\nu-\overline{\sigma}_\nu\sigma_\mu).
\end{eqnarray}
In terms of the superspace covariant derivatives (\ref{Dsdef}) and the matrix model parameters (\ref{superconn}),
the $(W,\overline{W})$ superfields take the form
\begin{align}
W_{\alpha a} &=\frac{i}{4}\overline{D}_{\dot{\alpha}} \overline{D}^{\dot{\alpha}}\mathcal{M}_{\alpha a}, \label{Weq}\\
\nonumber
\overline{W}_{\dot{\alpha}a}&=\frac{i}{4}\epsilon_{\dot{\alpha}\dot{\beta}}
\Big(\epsilon_{abc}\epsilon^{\alpha\beta}\overline{D}^{\dot{\beta}}\mathcal{M}_{\beta b}\mathcal{M}_{\alpha c}
+D^\alpha\overline{D}^{\dot{\beta}}\mathcal{M}_{\alpha a}-\frac{3}{2}(\overline{\sigma}^0)^{\dot{\beta}\alpha}\mathcal{M}_{\alpha a}
-\frac{1}{2}(\overline{\sigma}^0)^{\dot{\beta}\alpha}(\overline{\sigma}^0)^{\dot{\gamma}\beta}\{\overline{D}_{\dot{\gamma}},D_\beta\}\mathcal{M}_{\alpha a}\Big).
\end{align}
This way of writing the $\mathcal{F}_{\mu A}$'s makes it clear that what is known as the reality constraint,
\begin{equation}
\mathcal{F}_{\mu\nu}^\dagger=\mathcal{F}_{\mu\nu}, \label{realcons}
\end{equation}
entails precisely the desired uniqueness of (\ref{Mmua}), as it implies
\begin{equation}
\mathcal{M}_{0a}^\dagger=\mathcal{M}_{0a}+\textrm{gauge transformation}, \qquad \mathcal{M}_{ia}^\dagger=\mathcal{M}_{ia}.
\end{equation}
Notice that (\ref{realcons}) also relates the otherwise independent $(W,\overline{W})$ superfields.
They are now subject to satisfy $W_{\alpha a}^\dagger=\overline{W}_{\dot{\alpha}a}+\textrm{gauge transformation}$.
The above implementation of the constraints (\ref{reppcstr}) and (\ref{realcons}) yields the correct number of degrees of freedom
on the matrix model gauge superpotentials $\mathcal{M}_{Aa}$.

As already stated, it is convenient to express superfields, and in particular, $W$, as an expansion in the fermionic variables
$(\theta^\alpha,\overline{\theta}^{\dot{\alpha}})$.
By construction, the coefficients of the different powers of $\theta^\alpha$ and $\overline{\theta}^{\dot{\alpha}}$ will be matrices depending only on time.
These play the role of fields in a supermultiplet. The transformation of $W$ under translations $X_A$ on the superspace induces the supersymmetry transformations of the fields.
Notice however that $W$ in (\ref{Weq}), is gauge-covariant, and hence its expanded form will be gauge-dependent. We choose to work in the Wess-Zumino gauge.
In this gauge, only the physical (matrix model reduced) fields in the supermultiplet are non-vanishing and so the degrees of freedom are manifest.
The choice may be regarded as analogous to setting the Coulomb gauge in electrodynamics.
Altogether, we get
\begin{align}
W_{\alpha a}=-i\lambda_{\alpha a}+\theta_\alpha D_a-\frac{1}{2}(\sigma^\mu)_{\alpha\dot{\alpha}}(\overline{\sigma}^\nu)^{\dot{\alpha}\beta}\theta_\beta F_{\mu\nu}^a
+\theta^\beta\theta_\beta\left((\sigma^0)_{\alpha\dot{\alpha}}\dot{\overline{\lambda}}_{\dot{\beta}a}
+i(\sigma^0)_{\alpha\dot{\alpha}}\overline{\lambda}_{\dot{\beta}a}
+\epsilon_{abc}(\sigma^\mu)_{\alpha\dot{\alpha}}M_{\mu b}\overline{\lambda}_{\dot{\beta}c}\right)\epsilon^{\dot{\beta}\dot{\alpha}}, \label{WinWZ}
\end{align}
where the field strengths are given by
\begin{equation}
F_{0i}^a=\dot{M}_{ia}+\epsilon_{abc}M_{0b}M_{ic},
\qquad F_{ij}^a=-\epsilon_{ijk}M_{ka}+\epsilon_{abc}M_{ib}M_{jc}.
\end{equation}
Observe the different character of $\epsilon_{ijk}$ and $\epsilon_{abc}$ here:
the former signals that the bosonic spatial space $\mathbf{R}^3$  has been compactified to $S^3$,
while the latter captures the structure constants of the underlying $SU(2)$ gauge group.
It will be immediately relevant to also point out that $W$ is a chiral superfield, see (\ref{Weq}).

Finally, the matrix model action for $\mathcal{N}=1$,
$SU(2)$ super-Yang-Mills theory is derived by integrating the superfield Lagrangian over superspace.
In this case, the square of (\ref{WinWZ}) yields the Lagrangian of the theory.
Explicitly,
\begin{eqnarray}
S=\int_{S^3\times\mathbb{R}}d^4x\,
\mathcal{L}, \qquad
\mathcal{L}=\frac{1}{16}\int d^2\theta\,\,\epsilon^{\alpha\beta}W_{\beta a} W_{\alpha a}. \label{smmaction}
\end{eqnarray}
The Lagrangian for the matrix model can be written in a compact way as follows:
\begin{equation}
\mathcal{L}=-\frac{1}{4g^2}F_{\mu\nu}^aF^{a\mu\nu}
-\frac{i}{g^2}\overline{\lambda}^a_{\dot{\alpha}}(\overline{\sigma}^\mu)^{\dot{\alpha}\alpha} (\mathcal{D}_\mu\lambda_\alpha)^a
-\frac{1}{g^2}\overline{\lambda}^a_{\dot{\alpha}}(\overline{\sigma}^0)^{\dot{\alpha}\alpha}\lambda^a_\alpha+\frac{1}{2g^2}D^aD^a,
\end{equation}
where $g$ is the gauge coupling constant and the $\mathcal{D}_\mu$ are the gauge-covariant derivatives:
\begin{eqnarray}
(\mathcal{D}_0 f)^a=\partial_0 f^a+\epsilon_{abc}M_{0b}f^c, \quad\quad
(\mathcal{D}_i f)^a=\epsilon_{abc}M_{ib}f^c.
\end{eqnarray}
It can be readily seen that the field $D^a$ has no kinetic term.
It is an auxiliary field that vanishes on shell, while ensuring that the number of bosonic and fermionic components match off shell.

Until this point, we have chosen the radius of the spatial $S^3$ to be one,
but it is straightforward to rewrite our equations for arbitrary radius $\rho$ by a simple dimensional analysis.
The Lagrangian density then becomes
\begin{equation}
\label{N1Lagrangian}
\mathcal{L}=-\frac{1}{4g^2}F_{\mu\nu}^aF^{a\mu\nu}
-\frac{i}{g^2}\overline{\lambda}^a_{\dot{\alpha}}(\overline{\sigma}^\mu)^{\dot{\alpha}\alpha} (\mathcal{D}_\mu\lambda_\alpha)^a
-\frac{1}{g^2\rho}\overline{\lambda}^a_{\dot{\alpha}}(\overline{\sigma}^0)^{\dot{\alpha}\alpha}\lambda^a_\alpha+\frac{1}{2g^2}D^aD^a,
\end{equation}
with the field strength being modified as
\begin{equation}
F_{0i}^a=\dot{M}_{ia}+\epsilon_{abc}M_{0b}M_{ic},
\qquad F_{ij}^a=-\frac{1}{\rho}\epsilon_{ijk}M_{ka}+\epsilon_{abc}M_{ib}M_{jc}.
\end{equation}
The action corresponding to (\ref{N1Lagrangian}) is invariant under the following supersymmetry transformations:
\begin{align}
\nonumber
\delta M_{\mu a}&=i(\overline{\zeta}_{\dot{\alpha}}(\overline{\sigma}_{\mu})^{\dot{\alpha}\alpha}\lambda^a_{\alpha}
-\overline{\lambda}^a_{\dot{\alpha}}(\overline{\sigma}_\mu)^{\dot{\alpha}\alpha}\zeta_\alpha), \qquad\quad\,\,\,
\delta\lambda^a_\alpha=\frac{1}{2}(\sigma^{\mu\nu})_\alpha{}^\beta\zeta_\beta F_{\mu\nu}^a+i\zeta_\alpha D^a, \nonumber \\
\delta D^a&=\overline{\zeta}_{\dot{\alpha}}(\overline{\sigma}^{\mu})^{\dot{\alpha}\alpha}(\mathcal{D}_\mu\lambda_\alpha)^a
+(\mathcal{D}_\mu\overline{\lambda}_{\dot{\alpha}})^a(\overline{\sigma}^\mu)^{\dot{\alpha}\alpha}\zeta_\alpha
-\frac{i}{\rho}(\overline{\zeta}_{\dot{\alpha}}(\overline{\sigma}^0)^{\dot{\alpha}\alpha}\lambda^a_\alpha
-\overline{\lambda}^a_{\dot{\alpha}}(\overline{\sigma}^0)^{\dot{\alpha}\alpha}\zeta_\alpha), \label{Mvariation}
\end{align}
where $(\zeta^\alpha,\overline{\zeta}_{\dot{\alpha}})$ are the supersymmetry (anticommuting) parameters depending only on time.
Compatibility with supersymmetry then requires $\zeta$ to be a constant: $\partial_0\zeta=0$.
Clearly, the supersymmetry transformation of $\lambda^a_\alpha$ implies
$\delta\overline{\lambda}^a_{\dot{\alpha}}=
\frac{1}{2}\overline{\zeta}_{\dot{\beta}}(\overline{\sigma}^{\mu\nu})^{\dot{\beta}}{}_{\dot{\alpha}}F_{\nu\mu}^a
-i\overline{\zeta}_{\dot{\alpha}} D^a$.
Corresponding to the above supersymmetry, by Noether's theorem, there is a conserved supercharge $Q$.
This is computed to be
\begin{equation}
\label{charge}
Q=Q^\alpha \zeta_\alpha+\overline{\zeta}_{\dot{\alpha}}\overline{Q}^{\dot{\alpha}}, \qquad
Q^\alpha=\frac{i}{2g^2}\overline{\lambda}^a_{\dot{\alpha}}(\overline{\sigma}^0)^{\dot{\alpha}\beta}(\sigma^{\mu\nu})_\beta{}^\alpha F_{\mu\nu}^a, \qquad 
\overline{Q}^{\dot{\alpha}}=\frac{i}{2g^2}(\overline{\sigma}^{\mu\nu})^{\dot{\alpha}}{}_{\dot{\beta}}(\overline{\sigma}^0)^{\dot{\beta}\alpha} \lambda^a_\alpha F_{\mu\nu}^a.
\end{equation}

\subsection{The Hamiltonian}
\label{sechamiltonian}

The central object of study in supersymmetric quantum mechanics is the Hamiltonian.
Accordingly, in the following we make use of the equivalence between the Lagrangian and Hamiltonian formalisms~\cite{LagHam}
to obtain all relevant quantities of the just derived matrix model in the latter picture.
This will enable us to investigate our model's quantum phase structure in section \ref{secBornOpp}.
Henceforth, we shall omit contracted spinorial indices, so as to abbreviate the notation.
 
As a first step, we compute the conjugate momenta to $M_{ia}$ and $\lambda_a$ in (\ref{N1Lagrangian}):
\begin{equation}
\label{momenta}
\Pi_{ia}\equiv\frac{\partial \mathcal{L}}{\partial \dot{M}^{ia}}=\frac{1}{g^2}F_{0i}^a, \qquad
\Pi^{\alpha a}\equiv\frac{\partial \mathcal{L}}{\partial \dot{\lambda}^a_\alpha}=
-\frac{i}{g^2}(\overline{\lambda}^a\overline{\sigma}^0)^\alpha.
\end{equation}
Observe that $M_{0a}$ is non-dynamical: the Lagrangian does not depend on its time derivative $\dot{M}_{0a}$.
For this reason, its conjugate momentum vanishes\footnote{More precisely, this vanishing is a primary constraint.
Therefore, one should really write $\Pi_{0a}\approx0$.
Here, $\approx$ denotes a so-called weak equality, which only holds true in the subspace of the parameter space (known as the constraint surface)
that the constraint itself defines.}, $\Pi_{0a}=0$,
and so $M_{0a}$ plays the role of a Lagrange multiplier.
The only non-vanishing (anti)commutation relations among the matrix model fields and momenta can be easily verified to be of the canonical form:
\begin{align}
\left[M_{ia},\Pi_{jb}\right]=i\delta_{ij}\delta_{ab}, \quad\quad
\{\lambda^a_\alpha,\Pi^{b \beta}\}=i\delta_{ab}\delta_\alpha{}^{\beta}. \label{fieldmom}
\end{align}
Using (\ref{N1Lagrangian}) and (\ref{momenta}), it is a matter of straightforward algebra to calculate the matrix model Hamiltonian.
This is given by
\begin{equation}
H'=H+M_{0a}G_a, \qquad
H=\frac{g^2}{2}\Pi_{ia}\Pi_{ia}+\frac{1}{4g^2}F_{ij}^a F_{ij}^a
+\frac{1}{g^2\rho}\overline{\lambda}^a\overline{\sigma}^0\lambda^a+\frac{i}{g^2}\epsilon_{abc}\overline{\lambda}^a\overline{\sigma}^i\lambda^c M_{ib},
\label{hamonshell}
\end{equation}
with $H$ the on shell part of the Hamiltonian and $G_a$ standing for the Gauss law operator
\begin{equation}
\label{defGa}
G_a\equiv \epsilon_{abc}(\Pi_{ib}M_{ic}-\frac{i}{g^2}\overline{\lambda}^b\overline{\sigma}^0\lambda^c).
\end{equation}
The above operator generates infinitesimal color transformations and so satisfies an $SU(2)$ algebra: $[G_a,G_b]=-i\epsilon_{abc}G_c$.
We stress that the Gauss law operator vanishes when acting on physical states.
This will be relevant later on.

In terms of the momenta, the conserved supercharge's components in (\ref{charge}) become
\begin{eqnarray}
Q^\alpha= \frac{1}{g^2}(\overline{\lambda}^a\overline{\sigma}^i)^\alpha\Big(ig^2\Pi_{ia}+\frac{1}{2}\epsilon_{ijk}F_{jk}^a\Big),\qquad
\overline{Q}^{\dot{\alpha}}=\frac{1}{g^2}(\overline{\sigma}^i\lambda^a)^{\dot{\alpha}}\Big(-ig^2\Pi_{ia}+\frac{1}{2}\epsilon_{ijk}F_{jk}^a\Big).
\label{QQ}
\end{eqnarray}
It is easy to check that $Q$ forms a field representation of supersymmetry:
\begin{equation}
[Q,M_{ia}]=i\delta M_{ia}, \qquad 
[Q,\lambda^a_\alpha]=i\delta\lambda^a_\alpha,
\end{equation}
with $(\delta M_{ia},\delta\lambda^a_\alpha)$ as given in (\ref{Mvariation}).
Straightforward yet tedious algebra allows one to write the non-trivial (anti)commutator of the algebra among the supercharge's components as
\begin{equation}
\{Q^\alpha,\overline{Q}^{\dot{\beta}}\}=
-(\overline{\sigma}^0)^{\dot{\beta}\alpha}(2H+R)+2(\overline{\sigma}^{i})^{\dot{\beta}\alpha}(G_a M_{ia}+J^i),
\label{anticomm}
\end{equation}
where $J^i$ is the angular momentum operator generating rotations in the spatial $S^3$ and $R$ is the $R$-parity operator.
Explicitly,
\begin{equation}
\label{JRops}
J^i= \epsilon_{ijk}\Pi_{ja}M_{ka}+\frac{1}{2g^2}\overline{\lambda}^a\overline{\sigma}^i\lambda^a, \qquad
R=9+\frac{1}{g^2}\overline{\lambda}^a\overline{\sigma}^0\lambda^a.
\end{equation}
Thus the matrix model reproduces the known $R$-charges of the $\mathcal{N}=1$ super-Yang-Mills gauge multiplet.
In our conventions, this means that $M_{ia}$ is neutral, while $\lambda^a_\alpha$ is R-even:
$[R,M_{ia}]=0$ and $[R,\lambda_\alpha^a]=\lambda_\alpha^a$.
The other commutation relations required to describe the full superalgebra are
\begin{align}
[Q^\alpha,J_i]=\frac{1}{2}(Q\sigma^0\bar{\sigma}^i)^\alpha, \qquad
[Q^\alpha,R]=Q^\alpha, \qquad
[Q^\alpha,G_a]=0, \qquad
[Q^\alpha,H]=(\bar{\lambda}^a\bar{\sigma}^0)^\alpha G_a.
\label{othercomm}
\end{align}
Notice that the first commutator indicates that $Q$ transforms as a spin-$\frac{1}{2}$ operator.
Consequently, $Q$ has $R$-charge equal to one, as shown in the second commutator.
Since $G_a$ vanishes on the space of physical states, in this space of color-singlets, the Hamiltonian commutes with the supercharge in the physical Hilbert space 
--- see the last commutator of (\ref{othercomm}).
It follows then that degenerate eigenstates of $H$ organize themselves into supersymmetry multiplets.

\section{Born-Oppenheimer quantization of the supersymmetric matrix model}
\label{secBornOpp}

The Hamiltonian (\ref{hamonshell}) governs the dynamics of the gauge fields $M_{ia}$ and their superpartners $\lambda_\alpha^a$.
When the coupling constant $g$ is small, we observe that the kinetic term for the gauge fields is suppressed with respect to that of the fermions. In this weak coupling limit, it is suitable to quantize the system in the Born-Oppenheimer approximation,
as was argued in~\cite{Pandey:2016hat}, where the general framework of~\cite{BKLB} was suitably adapted to the matrix model case
in the presence of fundamental fermions.
In brief, the modern treatment of the said approximation consists on viewing the fermions as ``fast'' degrees of freedom
and quantizing them in the background of the (comparatively) ``slow'' gauge fields.
Then, the effective dynamics of the gauge bosons induced by the fermions is determined.
We begin this section by providing the details of this procedure.
Afterwards, we proceed to its implementation in sections \ref{sec31} and \ref{sec32}.

For notational convenience, we abandon the use of dotted spinor indices from this point onwards
and understand $\bar{\lambda}\equiv\lambda^\dagger$.
Paralleling the discussion in~\cite{Pandey:2016hat}, we begin by rewriting our on shell Hamiltonian in (\ref{hamonshell})
as a sum of its gauge and fermionic pieces,
\begin{align}
H=H_{YM}+H_f, \qquad
H_{YM} = \frac{g^2}{2}\Pi_{ia}\Pi_{ia}+\frac{1}{4g^2}F_{ij}^a F_{ij}^a, \qquad
H_f = -\frac{1}{g^2} \big(\frac{1}{\rho}\lambda^\dagger_{\alpha a}\lambda_{\alpha a}
+\frac{1}{2}(T_b)_{ac}\lambda^\dagger_{\alpha a} (\overline{\sigma}^i)_{\alpha \gamma}\lambda_{\gamma c} M_{ib} \big),
\label{onshellH}
\end{align}
with $(T_a)_{bc}\equiv-i\epsilon_{abc}$ the generators of gauge transformations.
We denote as $\mathcal{H}_{tot}$ the Hilbert space of physical states of $H$. 

If $g$ is sufficiently small, the fermion dynamics is much faster compared to the gauge dynamics and can be quantized separately,
treating the gauge degrees of freedom as a slow moving background field.
In this context, $\mathcal{H}_{tot}$ can be split into the direct product of the Hilbert spaces of the fast and slow degrees of freedom:
$\mathcal{H}_{tot}=\mathcal{H}_{slow}\otimes\mathcal{H}_{fast}$.
We first construct $\mathcal{H}_{fast}$ from the set of eigenstates of the fermionic Hamiltonian $H_f$,
obtained by treating the gauge field variables $M_{ia}$ as a background field and solving the eigenvalue equation
\begin{equation}
\label{eigenhf}
H_f(M)|n(M)\rangle=\epsilon_n(M)|n(M)\rangle,
\end{equation}
with $n\in\mathbb{N}\cup\{0\}$ labeling the energy levels.
A suitable choice for a complete set of basis vectors in $\mathcal{H}_{tot}$ is then given by the generalized eigenvectors
\begin{equation}
\label{MnMstate}
|M,n(M)\rangle\equiv|M\rangle\widetilde{\otimes}|n(M)\rangle,
\end{equation}
where $|M\rangle$ are the bosonic ``position'' vectors,
i.e.~eigenstates of the operator $M_{ia}$ that label the points in the (matrix model reduced) configuration space of Yang-Mills.
Note that $\widetilde{\otimes}$ indicates that the right-hand side of (\ref{MnMstate}) is not an ordinary tensor product
but rather a ``twisted'' direct product, since $|n(M)\rangle$ depends on the gauge fields $M_{ia}$.

Let $|\psi^E\rangle$ denote an eigenstate of the on shell Hamiltonian $H$ with eigenvalue $E$:
\begin{equation}
H|\psi^E\rangle=E|\psi^E\rangle.
\label{psiEdef}
\end{equation}
This energy eigenstate can be expanded in the basis (\ref{MnMstate}) as
\begin{equation}
|\psi^E\rangle= \int dM^\prime \sum_n |M^\prime,n(M^\prime)\rangle\, \psi^E_n(M^\prime), \qquad
\psi^E_n(M^\prime)\equiv \langle M^\prime,n(M^\prime)|\psi^E\rangle.
\end{equation}
Here, $\psi_n^E(M)$ can be thought of as the slow part of the wavefunction $|\psi^E\rangle$.
It satisfies the Schr\"{o}dinger equation
\begin{equation}
\sum_m \Big[\frac{g^2}{2}\sum_l (-i\delta^{nl}\partial_{ia}-A_{ia}^{nl})(-i\delta^{lm}\partial_{ia}-A_{ia}^{lm})
+\delta^{nm}\Big(\frac{1}{4g^2}F_{ij}^aF_{ij}^a + 
\epsilon_n(M)\Big)\Big]\psi_m^E(M)=E\psi_n^E(M), 
\label{slowSchrod}
\end{equation}
with $A_{ia}^{mn}\equiv i\langle n(M)|\partial_{ia}|m(M)\rangle$.
The above follows from taking the inner product on both sides of (\ref{psiEdef}) with the basis states defined in (\ref{MnMstate})
and working through.
The indices $l$, $m$ and $n$ run over all fermion energy levels.

Henceforth, we shall focus on the situation where a single fermion occupies the ground state.
To elaborate, we think of the matrix model as describing the regime of the field theory with a large baryon chemical potential
---as already noted in the introduction \ref{intro}.
In this regime, the quarks form a weakly interacting Fermi liquid (i.e.~a Fermi sea)
and the fermion vacuum can be thought of as the Fermi surface, with only a finite number of energy levels available near it.
We are thus interested in examining the effective gauge dynamics induced by a single fermion
excited to occupy the lowest energy level available near the Fermi surface.
More generally, we would investigate the case of a few fermions occupying the lowest available energy levels.
However and as we shall work out in details in the next section \ref{sec31},
the multi-fermion situation can be easily defined in terms of the single-fermion case.

Taking into account fermions that only occupy their ground state amounts to restricting to $n=0=m$ in (\ref{slowSchrod}).
In general, the fermionic ground state may be degenerate.
We label this degeneracy with Greek letters $(\alpha,\beta, \ldots)$, which take values from $1$ to $g_0$,
the degeneracy of the ground state.
In this case, the slow wavefunction $\psi^E_\alpha(M)$ satisfies
\begin{equation}
H_{eff}^{\alpha\beta}\psi^E_{\beta}=E\psi^E_\alpha, \qquad
H_{eff}^{\alpha\beta}=-\frac{g^2}{2}\mathcal{D}^{\alpha\gamma}_{ia}\mathcal{D}^{\gamma\beta}_{ia}+\delta^{\alpha\beta}
\Big(\frac{1}{4g^2} F_{ij}^aF_{ij}^a+\epsilon_0(M)+\frac{g^2}{2}\Phi(M)\Big).
\label{Heff}
\end{equation}
Here, $\mathcal{D}$ is the covariant derivative, whose explicit form is 
\begin{equation}
\mathcal{D}^{\alpha\beta}_{ia}=\delta^{\alpha\beta}\partial_{ia}-i \mathcal{A}^{\alpha\beta}_{ia},
\label{covder}
\end{equation}
with $\mathcal{A}^{\alpha\beta}_{ia}$ the vector potential induced by the fermion in the effective gauge dynamics:
\begin{equation}
\mathcal{A}^{\alpha\beta}_{ia}\equiv i\langle 0(M),\alpha|\partial_{ia}|0(M),\beta\rangle.
\label{vecpot}
\end{equation}
Notice that the fermion induces an additional effective scalar potential $\Phi$ for the slow degrees of freedom $M_{ia}$.
The $\Phi$ can be expressed in terms of the projector $\mathbb{P}_0$ to the ground state
and $\mathbb{Q}_0\equiv\mathbf{1}-\mathbb{P}_0$ as~\cite{Zanardi}
\begin{equation}
\Phi=\sum_{l\neq0}A_{ia}^{0l}A^{l0}_{ia}=\frac{1}{g_0}{\rm Tr}\left(\mathbb{P}_0\partial_{ia}H_f
\frac{\mathbb{Q}_0}{(H-\epsilon_0)^2}\partial_{ia}H_f \mathbb{P}_0\right).
\label{phiasP}
\end{equation}
Both $\mathcal{A}_{ia}^{\alpha\beta}$ and $\Phi$ are best understood in the context of quantum adiabatic transport.
In the first step of our approximation,
we need to quantize the fermions in the background of the slowly varying gauge fields $M_{ia}$, see (\ref{eigenhf}).
The $M_{ia}$'s act as an adiabatic parameters on which the Hamiltonian $H_f$ and its spectrum have functional dependence.
The induced vector potential $\mathcal{A}_{ia}^{\alpha\beta}$ in (\ref{vecpot})
is simply the Berry connection associated with the ground state of $H_f$,
while the effective scalar potential $\Phi$ is the trace of the quantum metric tensor~\cite{berry}.
The latter acts as a measure of the ``distance" between two states corresponding to the same energy level (the ground state in this case),
but separated in the parameter space.

Having set up the Born-Oppenheimer quantization for the on-shell part of the Hamiltonian (\ref{hamonshell}),
we now turn to its off shell piece.
Following a procedure analogous to that which allowed us to obtain the effective Hamiltonian (\ref{Heff}) from (\ref{psiEdef}),
the Gauss' law constraint $G_a|\psi^E\rangle=0$ results into a modified Gauss' law generator $\mathcal{G}_a^{\alpha\beta}$.
This can be worked out to be
\begin{align}
\label{effga}
\mathcal{G}_a^{\alpha\beta}= i\delta^{\alpha \beta} \epsilon_{abc}M_{ib}\partial_{ic}+\langle 0(M),\alpha|G_a|0(M),\beta\rangle.
\end{align}
We observe that, since $H_f$ is gauge-invariant, its eigenstates must also be annihilated by Gauss' law generators:
$G_a|n(M)\rangle=0$, for any eigenstate $|n(M)\rangle$.
In this case, the effective Gauss' law operator reduces to the first term in (\ref{effga})
and it is easy to verify that such $\mathcal{G}_a^{\alpha\beta}$'s satisfy an $SU(2)$ commutation relation:
$[\mathcal{G}_a,\mathcal{G}_b]^{\alpha\beta}=-i\epsilon_{abc}\mathcal{G}_c^{\alpha\beta}$.

Similarly, we obtain expressions for the effective angular momentum and effective $R$-charge operator
acting on the Hilbert space of $H_{eff}$:
\begin{equation}
\mathcal{J}_i^{\alpha\beta}=i\epsilon_{ijk}M_{ja}\mathcal{D}_{ka}^{\alpha\beta}-\frac{1}{2g^2}\langle 0(M),\alpha|\lambda^\dagger(\sigma_i\otimes\mathbf{1})\lambda|0(M),\beta\rangle, \qquad
\mathcal{R}^{\alpha\beta}=(9-r_0)\delta^{\alpha\beta},
\label{JReff}
\end{equation}
where $r_0$ counts the number of fermions in the ground state $|0(M)\rangle$.

To summarize, the Born-Oppenheimer approximation procedure involves first calculating the fermionic energy spectrum
by treating the gauge variables $M_{ia}$ as a background field; namely, we should solve (\ref{eigenhf}).
We then focus on the ground state energy of $H_f$ and its corresponding (possibly degenerate) eigenstate.
The effective gauge dynamics thereby induced should be determined via (\ref{Heff}), (\ref{effga}) and (\ref{JReff}).

\subsection{The fermionic spectrum}
\label{sec31}

We now proceed with the first step of the Born-Oppenheimer quantization procedure, i.e.~we turn to solving (\ref{eigenhf}).
To simplify notation, we denote by capital Latin letters the collective color and spin indices: $A\equiv (a,\alpha)$, $A=1,...,6$.
Then, we can concisely rewrite $H_f$ in (\ref{onshellH}) as
\begin{equation}
H_f(M)=-\lambda^\dagger_A (H_f)_{AB}\lambda_B, \qquad (H_f)_{AB}=(-\frac{1}{\rho}\mathbf{1}-M_{ic}T_c \otimes \sigma^i)_{AB},
\end{equation}
where $\mathbf{1}\equiv 1\otimes 1$.
Since the above Hamiltonian commutes with the fermion number operator $\lambda^\dagger_A\lambda_A$,
its eigenstates can be arranged according to their fermion number: $|r,n(M)\rangle$.
For every fermion number $r=0,1,\ldots,6$; $n$ runs over all possible $r$-fermion eigenstates.
The fermionic vacuum $|0\rangle$ is non-degenerate and has zero energy.
It is not difficult to see that the normalized one-fermion eigenstates are of the form
\begin{equation}
|1,n\rangle =\frac{1}{g} C^{1,n}_{A}\lambda^\dagger_{A}|0\rangle, \qquad \textrm{such that}\quad
g^2(H_f)_{AB}C^{1,n}_B=\epsilon_{1,n} C^{1,n}_A.
\label{single}
\end{equation}
The above is just a (suitably normalized) linear combination of one-fermion states with fixed spin and color.
Then, states with higher fermion number can be constructed by placing fermions in different spin-color single-fermion energy levels:
\begin{align}
|r>1,n\rangle=\frac{1}{g^r\sqrt{r!}}C^{r,n}_{A_1\ldots  A_r}\lambda^\dagger_{A_1}\ldots\lambda^\dagger_{A_r}|0\rangle.
\end{align}
Because any two $\lambda^\dagger$'s anticommute, the $C^{r,n}$'s are antisymmetric under the exchange of any two pairs of indices $A_i$ and $A_j$,
with $i\neq j$.
It can be easily verified that
\begin{equation}
C^{r,n}_{A_1\ldots A_r}=C^{1,n_1}_{\{A_1}C^{1,n_2}_{A_2}\ldots C^{1,n_r}_{A_r\}},
\end{equation}
with the energy of the corresponding eigenstate being
\begin{equation}
\epsilon_{r,n}=\sum_{i=1}^r \epsilon_{1,n_i}.
\end{equation}

To sum up, single-fermion states can be constructed by evaluating the eigenvectors $C_A^{1,n}$ of $(H_f)_{AB}$
and taking a linear combination, see (\ref{single}).
States with higher fermion number can then be constructed by placing fermions in different single-fermion energy levels according
to the Pauli exclusion principle, with the maximum number of fermions that can be placed being six.
Therefore, it is sufficient to consider only single-fermion energy levels for our following discussion
---since every other result can be easily deduced from these.
For ease of notation, henceforth we omit the $1$ in the superscript of the single-fermion eigenstate: $C^n_A\equiv C^{1,n}_A$.

Ignoring the constant $-\frac{1}{\rho}\mathbf{1}$ piece of $(H_f)_{AB}$,
its characteristic polynomial $f(x)^2\equiv\textrm{det}\big(x\mathbf{I}-(H_f)_{AB}\big)$ is evaluated to be
$f(x)^2=(x^3-\textrm{Tr}(M^TM)x-2\det M)^2$.
We define the second-degree gauge-invariant function of $M$ as
\begin{equation}
\mathbf{g}_2\equiv \left(\frac{\text{Tr}M^TM}{3}\right)^{1/2}.
\end{equation}
Upon rescaling $x$ as $x \rightarrow x/\mathbf{g}_2$, the characteristic polynomial takes the simpler form
\begin{equation}
\label{lasteq}
f(x)^2=(x^3-3x-2\mathbf{g}_3)^2, \qquad \mathbf{g}_3\equiv\frac{\text{det }M}{(\mathbf{g}_2)^{3}}.
\end{equation}
We denote as $x_n$ its roots: $f(x_n)^2=0$.
The single-fermion energy eigenvalues follow from these, according to the relation
\begin{equation}
\epsilon_{1,n}=-1+\mathbf{g}_2 x_n.
\end{equation}
Note that $x_n$ and $\epsilon_{1,n}$ have functional dependence on the gauge-invariant functions of $M$,
given by $\mathbf{g}_2$ and $\mathbf{g}_3$.
These functions can be used as coordinates on the gauge configuration space.
While the explicit expression for the roots $x_n$ is cumbersome and not of much physical significance,
we can gain valuable insights into the fermion spectrum by analyzing the characteristic polynomial $f(x)^2$ itself.

Firstly, we observe that there are three doubly-degenerate energy levels, given by the roots of the cubic $f(x)$.
The double-degeneracy is a characteristic of adjoint fermions.
Indeed, given a single-fermion eigenstate of $H_f$ of the form
\begin{eqnarray}
\label{1fer}
|\psi^{1,n}\rangle =C^n_{A}(M)\lambda^\dagger_{A}|0\rangle
\end{eqnarray}
there is a degenerate single-fermion eigenstate
\begin{eqnarray}
|\chi^{1,n}\rangle=(\sigma_2{C^{n}}^*)_{A}\lambda^\dagger_{A}|0\rangle.
\label{timerev}
\end{eqnarray}
It is easy to see that the degenerate states $|\psi^{1,n}\rangle$ and $|\chi^{1,n}\rangle$ are related to each other via time-reversal.
Hence, the double-degeneracy of the single-fermion spectrum of adjoint fermions is a consequence of Kramers' theorem~\cite{Kram}.

\begin{figure}[hb]
\centering
\includegraphics[scale=1]{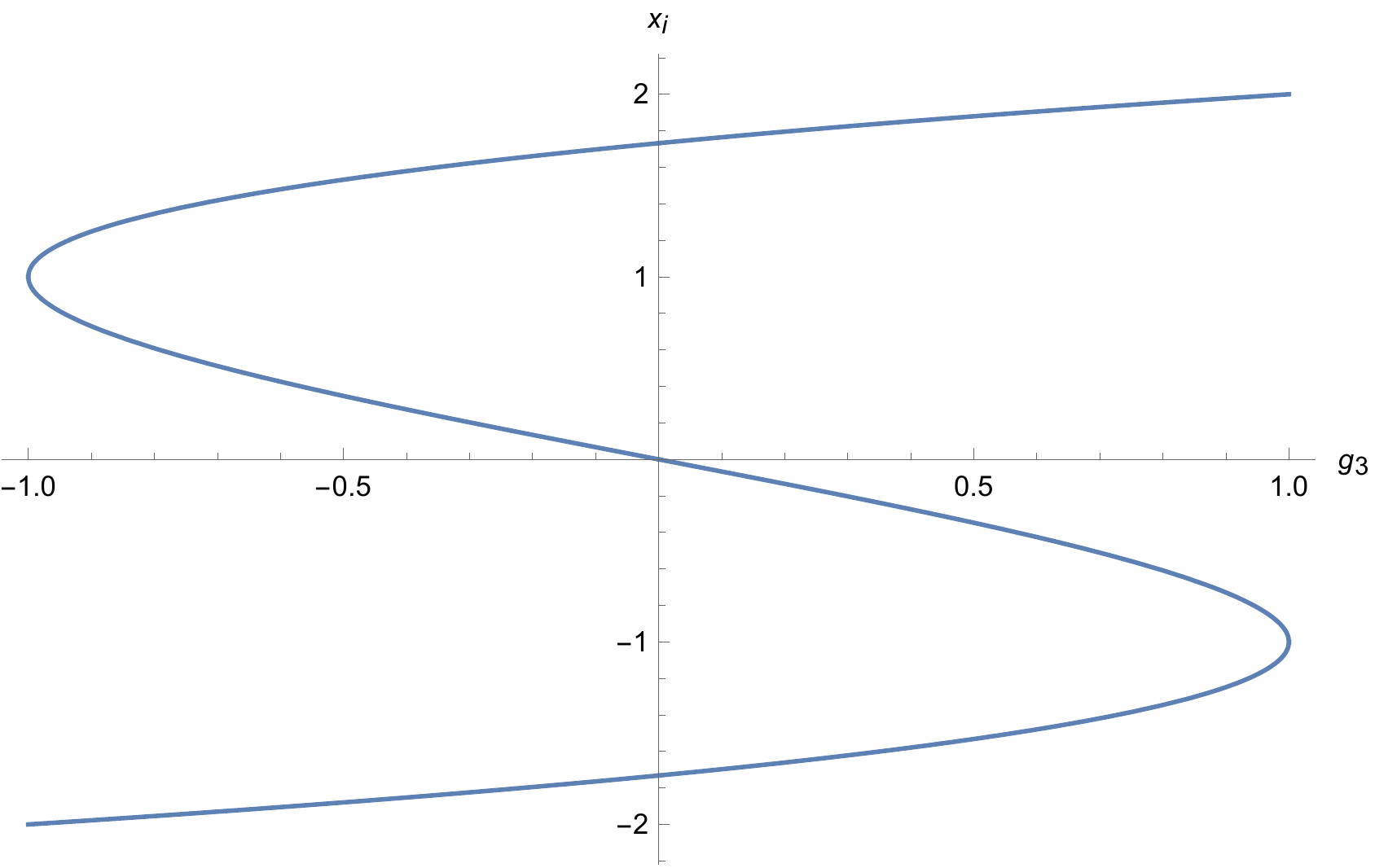}
\caption{Roots $x_i$ of the characteristic polynomial as a function of $\mathbf{g}_3$.}
\label{xvsg}
\end{figure}

Next, we examine the cubic polynomial $f(x)$.
Because $(H_f)_{AB}$ is a Hermitian matrix, $f(x)$ must have three real roots.
Since the leading term of $f(x)$ is cubic, $\lim_{x\to\pm\infty}f(x)=\pm\infty$.
It follows from both considerations that the curve $f(x)$ must intersect the $x$-axis three times and its local minimum must be negative or zero. We start by localizing the extrema of $f(x)$:
\begin{equation}
\frac{df}{dx}=0\implies x=\pm1.
\end{equation}
It is easy to check that $x=1$ is a minimum, while $x=-1$ is a maximum.
Thus, the condition for $f(x)$ to have all three real roots is
\begin{eqnarray}
f\Big|_{x=+1}\leq 0\implies |\mathbf{g}_3|\leq 1.
\label{fcubicmin}
\end{eqnarray}
The latter is a mathematical identity satisfied by any arbitrary real $3\times 3$ matrix $M$.
A plot of the roots of $f(x)$ against $\mathbf{g}_3$ in the allowed range is shown in figure \ref{xvsg}.

The inequality in (\ref{fcubicmin}) is saturated for $M= aR$, with  $a\in \mathbb{R}$ and $R\in SO(3)$.
In this case, the cubic polynomial reduces to
\begin{equation}
f(x)\Big|_{\mathbf{g}_3=1}=(x^3-3x-2)=(x-2)(x+1)^2,
\end{equation}
giving rise to the roots $x_1=x_2=-1$ and $x_3=2$.
At this corner of the configuration space, the ground state degeneracy changes from $2$ to $4$.

Recall that multiple-fermion energy levels can be deduced in a straightforward manner from the single-fermion energies.
For completeness, we provide the characteristic polynomials, roots and degeneracy of the complete fermionic spectrum in table \ref{table1}, starting from the vacuum state and all the way to the six-fermion state.
Remarkably, there is a fermion/hole duality in the spectrum:
the six-fermion state with all energy levels filled is equivalent to the vacuum, the five-fermion spectrum is equivalent to the single-fermion spectrum, and so on.
In particular, the three-fermion spectrum is self-dual.

\begin{table}[ht]
\centering
\begin{tabular}{|c|c|c|c|}
\hline 
Type & Roots & Degeneracy & Characteristic Polynomial \\ 
\hline 
0-fermion (vacuum) & 0 & 1 & $\emptyset$ \\ 
\hline 
& $x_1$ & 2 &  \\ 
1-fermion & $x_2$ & 2 & $(x^3-\textrm{Tr}(M^TM)x-2\textrm{det}M)^2 $\\
& $-(x_1+x_2)$ & 2 &  \\ 
\hline 
& $2x_1$ & 1 & \\
& $2x_2$ & 1 & \\
2-fermions & $-2(x_1+x_2)$ & 1 & $(x^3-4\textrm{Tr}(M^TM)x-16\textrm{det}M)\cdot$ \\ 
& $-x_1$ & 4 & $\cdot(x^3-\textrm{Tr}(M^TM)x+\textrm{det}M)^4$\\
& $-x_2$ & 4 & \\
& $x_1+x_2$ & 4 &  \\ 
\hline 
& $2x_1+x_2$ & 2 &  \\ 
& $-(2x_1+x_2)$ & 2 &  \\ 
& $x_1+2x_2$ & 2 & $x^8[x^6-6\textrm{Tr}(M^TM)x^4+(\textrm{Tr}(M^TM))^2x^2$ \\ 
3-fermions & $-(x_1+2x_2)$ & 2 & $-\{4(\textrm{Tr}(M^TM))^3-27(\textrm{det}M)^2\}^{1/2}]^2$ \\
& $x_1-x_2$ & 2 &  \\ 
& $-(x_1-x_2)$ & 2 &  \\ 
& 0& 8 &  \\  
\hline 
4-fermion = 2-hole & \multicolumn{3}{c|}{Negative of 2-fermion energies}  \\ 
\hline 
5-fermion = 1-hole & \multicolumn{3}{c|}{Negative of 1-fermion energies}  \\
\hline 
6-fermion & \multicolumn{3}{c|}{Same as vacuum:} \\ 
(completely filled) & \multicolumn{3}{c|}{$E$=0} \\
\hline
\end{tabular}
\caption{Roots, degeneracy and characteristic polynomial of the fermionic eigenstates of the supersymmetric matrix model
in the Born-Oppenheimer approximation, arranged according to their fermion number.}
\label{table1}
\end{table}

\subsection{Effective gauge dynamics}
\label{sec32}

After having determined the fermionic energy spectrum,
we proceed to examine the effective dynamics of the gauge degrees of freedom induced by the fermion occupying the ground state, or the lowest available energy level near the Fermi surface.
From (\ref{Heff}), it can be readily seen that this dynamics is governed by the effective potential
\begin{equation}
V_{eff}=\frac{1}{g^2}(V(M)+\epsilon_0)+\frac{g^2}{2}\Phi, \qquad V(M)\equiv\frac{1}{4}F_{ij}^aF_{ij}^a,
\end{equation}
where $\Phi$ is the induced scalar potential defined in (\ref{phiasP}).
The potential $V(M)$ and the ground state fermion energy $\epsilon_0$ are well-defined everywhere in the gauge configuration space.
However, $\Phi$ becomes singular whenever the ground state degeneracy changes.
We demonstrate this important point using the example of a single fermion occupying the ground state.

In this case, the scalar potential in the bulk of the gauge configuration space (i.e.~$\mathbf{g}_3<1$) is
\begin{equation}
\Phi_{bulk}=\frac{1}{9\mathbf{g}_2^2(1+x_1)^4}\left[ 7(1-x_1^2)^2(2+x_1^2)-2(1+2x_1^2)\left(1-\frac{\mathbf{g}_4}{3}\right) \right],
\qquad \mathbf{g}_4\equiv\frac{\text{Tr }(M^TM)^2 }{(\mathbf{g}_2)^2},
\label{phibulk}
\end{equation}
where $x_1$ stands for the lowest solution of the characteristic polynomial.
When the boundary point is approached from within the bulk (namely, $\mathbf{g}_3\to 1$ in the above equality),
the scalar potential shows a quadratic divergence:
\begin{equation}
\lim_{\mathbf{g}_3\to 1}\Phi_{bulk} = \frac{7}{12 \mathbf{g}_2^2}\frac{1}{(1+x_1)^2},
\end{equation}
since $\lim_{\mathbf{g}_3\to 1}\mathbf{g}_4= 3$ and $\lim_{\mathbf{g}_3\to 1}x_1 = -1$.
The effective potential $V_{eff}$ thus blows up as the boundary point is reached from the bulk.
To avoid inconsistencies, the wavefunctions in the domain of the effective Hamiltonian (\ref{Heff}) in the bulk must vanish at this boundary point.
However, if we instead restrict ourselves to the point $\mathbf{g}_3= 1$ and use the rank-4 projector to the ground state,
we obtain a well-defined scalar potential:
\begin{equation}
\Phi_{boundary}= \frac{7}{27\mathbf{g}_2^2}.
\end{equation}
This leads to an also well-defined $V_{eff}$ and nontrivial wavefunctions in the domain of the effect Hamiltonian at the boundary.
These cannot be consistently created out of a linear superposition of the bulk wavefunctions, since the latter have to vanish at the boundary.
It follows that the effective theory has two superselection sectors,
corresponding to the bulk $\mathbf{g}_3<1$ and to the boundary point $\mathbf{g}_3=1$.
These can be interpreted as two distinct phases.
As argued in~\cite{Pandey:2016hat}, the phase at the boundary is characterized by color-spin locked fermion condensates.

The above discussion can be readily generalized to states with higher fermion number.
As already stated, the fermionic state with $r$ fermions occupying the lowest available energy levels
is equivalent to a single fermion occupying the lowest $r$-fermion energy level.
So the induced effective potential can be computed using the corresponding information in table \ref{table1}.
In all cases, the singularity structure of $\Phi$ (and hence the two different phases)
can be easily identified from the degeneracy structure of the lowest root of the characteristic polynomial.

An important comment is in order here.
If instead of restricting to the ground state in (\ref{slowSchrod}) we took into account all the energy levels,
the induced effective potential would be zero, since $\mathbb{P}_0=\mathbf{1}$ in this situation.
Furthermore, the induced vector potential in (\ref{slowSchrod}) can be found to have zero curvature, rendering it pure gauge.
Thus, the induced gauge dynamics in this case would be trivial, corresponding to a situation without fermions.
This is expected, since the state with all six energy levels filled is equivalent to the vacuum,
due to the fermion/hole duality mentioned at the end of section \ref{sec31}. 

\section{Supersymmetry in the Born-Oppenheimer picture}
\label{sec4}

Throughout the analysis in section \ref{secBornOpp}, we have ignored the crucial point that the matrix model constructed is supersymmetric, while the Born-Oppenheimer approximation treats the gauge bosons (gluons) and the fermions (gluinos) differently.
As a result, we seem to have lost explicit supersymmetry upon quantizing the model.
To justify the approximation and its results, we need to recover supersymmetry in the effective theory.
This is the aim of the present section \ref{sec4}.

First, let us recast the relevant ideas of section \ref{secBornOpp}.
Before any approximation is made, the on shell Hamiltonian $H$ of the supersymmetric matrix model is that in (\ref{onshellH}).
Recall (\ref{psiEdef}), which defines $|\psi^E\rangle$ as an eigenstate of $H$ with energy $E$.
On the space of physical states $\mathcal{H}_{tot}$, $H$ commutes with the supercharges.
This implies that $Q_\alpha |\psi^E\rangle$ and $Q^\dagger_\alpha|\psi^E\rangle$ are also eigenstates of $H$ with eigenvalue $E$.
Additionally, $H$ commutes with the fermion-number operator, so its eigenstates have a fixed fermion number $r$.
Equation (\ref{slowSchrod}) relates the eigenfunctions of $H$ to all the eigenstates of the fermionic Hamiltonian $H_f$
with the same fermion number, through a modified Schr\"{o}dinger equation for the bosonic part of the wavefunction.
When we assume that only the ground states of the fermions contribute,
we obtain an effective Hamiltonian $H_{eff}^{(r)}$ in the $r$-fermion sector, given by (\ref{Heff}).

Next, we retrieve supersymmetry in our analysis.
We begin by considering the simple example of a purely bosonic eigenstate (with fermion number $r=0$), which we call $|\psi,0\rangle$.
It fulfills $H|\psi;0\rangle=E|\psi;0\rangle$, for some energy $E$.
In the Born-Oppenheimer approximation, the corresponding effective Hamiltonian for the slow degrees of freedom
satisfies the Schr\"{o}dinger equation
\begin{equation}
(H_{eff}^{(0)}) \psi^{(0)}=E\psi^{(0)},
\end{equation}
where the superscript $(0)$ denotes that we are in the zero-fermion sector.
The Fock vacuum $|0\rangle$ is unique, so there is no degeneracy and the dimension of $H_{eff}^{(0)}$ is one.
Since $Q^{\dagger}$ contains one $\lambda$ operator, it annihilates the purely bosonic state.
On the other hand, $Q$ contains one $\lambda^\dagger$ operator, so when it acts on the bosonic state it produces a single-fermion eigenstate with the same eigenvalue $E$:
\begin{equation}
Q_{\alpha} |\psi,0\rangle=|\psi,\alpha\rangle \implies H|\psi,\alpha\rangle=E|\psi,\alpha\rangle.
\label{QonState}
\end{equation}
Corresponding to $|\psi,\alpha\rangle$, there is an effective Hamiltonian in the single-fermion sector such that
\begin{equation}
(H_{eff}^{(1)})_{\rho\sigma} (\psi^{(1)}_\alpha)_{\sigma}=E(\psi^{(1)}_\alpha)_{\rho},
\end{equation}
where $\rho$ and $\sigma$ run over the degenerate ground states of the single-fermion sector.
Taking the inner product on both sides of (\ref{QonState}) with the basis vector $|M,n(M)\rangle$ and working through,
we get an effective supersymmetry charge which relates $\psi^{(0)}$ to $\psi^{(1)}$ as
\begin{equation}
(\mathcal{Q}_{\alpha}^{(0)})_{\rho} \psi^{(0)}=(\psi^{(1)}_\alpha)_{\rho}, \qquad(\mathcal{Q}_{\alpha}^{(0)})_{\rho}=(\overline{C}(M)^{\rho}_a \overline{\sigma}^i)_\alpha (-ig^2\partial_{ia}+\frac{1}{2}\epsilon_{ijk}F_{jk}^a).
\label{eq44}
\end{equation}
For each value of $\alpha $, the operator $\mathcal{Q}^{(0)}_\alpha$ is a rectangular $g_0\times 1$ matrix,
with $g_0$ the ground state degeneracy of the single-fermion spectrum.
Note that, since $\mathcal{Q}^{(0)}_\alpha$ is an operator with spin $\frac{1}{2}$,
the zero-fermion wavefunction $\psi^{(0)}$ gets related to a degenerate spin-$\frac{1}{2}$ doublet of states through supersymmetry.

Acting once more on the bosonic eigenstate with $Q_\beta$ (such that $\beta\neq\alpha$), a two-fermion state is obtained.
This yields an effective supersymmetry charge in the single-fermion sector,
that relates an eigenstate of $H_{eff}^{(1)}$ to an eigenstate of $H_{eff}^{(2)}$:
\begin{equation}
(\mathcal{Q}_{\beta}^{(1)})_{\rho^{(2)}\sigma}(\psi^{(1)}_\alpha)_{\sigma}=(\psi^{(2)}_{\alpha\beta})_{\rho^{(2)}},
\end{equation}
where $\rho^{(2)}$ runs over the degenerate two-fermion ground state.
Note that due to the anticommutation of the $Q$'s, $\psi_{\alpha\beta}=-\psi_{\beta\alpha}$, which implies there is only one such state.
Also, further action on this state with $Q$ leads to its annihilation.
The explicit form of $\mathcal{Q}^{(1)}$ can be worked out by noting that the two-fermion ground state is made up of two single-fermion states,
i.e.~$\rho^{(2)}=(\rho_1,\rho_2)$; so that $\rho^{(2)}$ runs over all such combinations.
We find that
\begin{equation}
(\mathcal{Q}_{\beta}^{(1)})_{\rho_1\rho_2,\sigma}=(\overline{C}_a^{\rho_1}\overline{\sigma}^i)_\beta(-ig^2\mathcal{D}_{ia}^{\rho_2 \sigma}+\frac{1}{2}\epsilon_{ijk}F_{jk}^a\delta^{\rho_2 \sigma})-(\rho_2\leftrightarrow\rho_1).
\end{equation}
Conversely, there is an operator $\mathcal{Q}^{(2)\dagger}_{\alpha}$ that takes the two-fermion state to any of the two single-fermion states.
Subsequent action of $\mathcal{Q}^{(1)\dagger}_\alpha$ takes the reached single-fermion state to the purely bosonic state.
These operators take the form
\begin{align}
(\mathcal{Q}_{\beta}^{(2)\dagger})_{\sigma,\rho_1\rho_2,}&=\left(\delta^{\sigma\rho_2}(\overline{\sigma}^iC_a^{\rho_1})_\beta-\delta^{\sigma\rho_1}(\overline{\sigma}^iC_a^{\rho_2})_\beta\right)(-ig^2\partial_{ia}+\frac{1}{2}\epsilon_{ijk}F_{jk}^a)-ig^2\left(\mathcal{D}^{\sigma\rho_2}(\overline{\sigma}^iC_a^{\rho_1})_\beta-\mathcal{D}^{\sigma\rho_1}(\overline{\sigma}^iC_a^{\rho_2})_\beta\right), \nonumber\\
(\mathcal{Q}_{\alpha}^{(1)\dagger})_\rho&=ig^2(\overline{\sigma}^i(\partial_{ia}C^\rho+C^\rho\partial_{ia})_{\alpha}+\frac{1}{2}\epsilon_{ijk}F_{jk}^a(\overline{\sigma}^i C^\rho)_\alpha.
\label{eq47}
\end{align}

In general, the full Hilbert space for the effective dynamics of the gluons can be expressed as a direct sum of sectors
\begin{equation}
\mathcal{H}=\bigoplus_r \mathcal{H}^{(r)},
\end{equation}
where $\mathcal{H}^{(r)}$ stands for the Hilbert space of the effective gauge dynamics
induced by $r$ fermions occupying the lowest energy levels.
In each of these sectors, there is an effective Hamiltonian $H_{eff}^{(r)}$ governing the gauge dynamics
and one can define effective supersymmetry charges that operate between pairs of Hilbert spaces:
\begin{align}
\mathcal{Q}_\alpha^{(r)}:\,\,\mathcal{H}^{(r)}\rightarrow \mathcal{H}^{(r+1)}, \qquad
\mathcal{Q}_\alpha^{(r)\dagger}:\,\,\mathcal{H}^{(r)}\rightarrow \mathcal{H}^{(r-1)}.
\end{align}
$\mathcal{Q}_\alpha^{(r)}$ takes us from the spectrum of $H_{eff}^{(r)}$ to that of $H_{eff}^{(r+1)}$.
Meanwhile, $\mathcal{Q}^{(r)\dagger}_\alpha$ takes us from the spectrum of $H_{eff}^{(r)}$ to that of $H_{eff}^{(r-1)}$:
\begin{align}
\mathcal{Q}_\alpha^{(r)}H_{eff}^{(r)}=H_{eff}^{(r+1)}\mathcal{Q}_\alpha^{(r)}, \qquad
\mathcal{Q}_\alpha^{(r)\dagger}H_{eff}^{(r)}=H_{eff}^{(r-1)}\mathcal{Q}_\alpha^{(r)\dagger}.
\label{eq410}
\end{align}
Since the $\mathcal{Q}$'s connect two distinct Hilbert spaces, they are represented by rectangular matrices.
The general expression for the effective supersymmetry charges is given by
\begin{equation}
(\mathcal{Q}_\alpha^{(r)})_{\Omega \rho}=\langle 0^{(r+1)},\Omega|Q|0^{(r)},\rho\rangle, \qquad
(\mathcal{Q}_\alpha^{(r+1)\dagger})_{ \rho\Omega}=\langle 0^{(r)},\rho|Q|0^{(r)},\Omega\rangle.
\end{equation}
Here, $|0^{(r)}\rangle$ denotes the lowest $r$-fermion energy level,
and $\rho$ and $\Omega$ run over the degenerate ground states of the $r$- and $(r+1)$-fermion sectors, respectively.
Thus, supersymmetry in the Born-Oppenheimer picture translates into a duality between Hilbert spaces of different effective Hamiltonians,
with the effective supercharges connecting the said spaces.

The supercharges satisfy an anticommutation relation:
\begin{equation}
\mathcal{Q_\alpha}^{(r+1)\dagger}\mathcal{Q_\beta}^{(r)}+\mathcal{Q_\alpha}^{(r-1)}\mathcal{Q_\beta}^{(r)\dagger}= \delta_{\alpha\beta}(2H_{eff}^{(r)}+\mathcal{R}^{(r)})-2(\sigma^i)_{\alpha\beta}(M_{ia}\mathcal{G}_a^{(r)}+\mathcal{J}_i^{(r)})-\Theta^{(r)}_{\alpha\beta},
\end{equation}
where $\Theta^{(r)}_{\alpha\beta}$ is given by
\begin{equation}
\Theta^{(r)}_{\alpha\beta}=\text{Tr}\left(\mathbb{P}_0^rQ^{\dagger}_\alpha
(1-\mathbb{P}_0^{r+1})Q_{\beta}\mathbb{P}_0^r\right)
+\text{Tr}\left(\mathbb{P}_0^rQ^{\dagger}_\alpha (1-\mathbb{P}_0^{r-1})Q_{\beta}\mathbb{P}_0^r\right).
\label{close}
\end{equation}
Here, $\mathcal{G}_a^{(r)},\mathcal{R}^{(r)}$ and $\mathcal{J}_i^{(r)}$
are the effective Gauss' law, $R$-charge operator and angular momentum in $\mathcal{H}^{(r)}$, respectively.
$\mathbb{P}_0^r$ is the projector to the ground state of the $r$-fermion sector.
The additional piece $\Theta^{(r)}_{\alpha\beta}$ arises because of our restriction on fermions to only occupy the lowest energy levels
and prevents the effective supersymmetry algebra from closing exactly.
The algebra would close if we took into account all the fermion energy levels: {$\mathbb{P}_0^{r}=\mathbb{I}$, for all $r$}.
However, that would describe trivial gauge dynamics {(as in the end of section \ref{sec32})}
and supersymmetry would be exact to begin with.

At the corner of the gauge configuration space described by $\mathbf{g}_3=1$,
the degeneracy of the $r$-fermion ground state changes for all $r=2,3,4,5$.
As a result, there is a different set of effective supercharges relating the effective Hamiltonians in this phase.
The two sets of effective supercharges {(in the bulk $\mathbf{g}_3<1$ and boundary $\mathbf{g}_3=1$)}
cannot be smoothly transformed into each other, since they have different dimensions.
Further, they satisfy different algebras; the projectors in (\ref{close}) having different ranks {in the distinct phases}.
Therefore, the effective supercharges exhibit a similar singular behavior as the effective scalar potential {in (\ref{phiasP})
and we say they are superselected}. 

In the beginning of this section \ref{sec4}, we identified one type of supersymmetry multiplet that definitely exists in the spectrum:
the one obtained by the action of the supercharge $\mathcal{Q}$ on purely bosonic states (or the fermion vacuum).
Correspondingly, there exists another multiplet that can be obtained by the action of $\mathcal{Q}^\dagger$
on six-fermion states (or the hole vacuum).
It can definitely be expected that there exists other multiplets;
these are obtained by the action of $\mathcal{Q}$ on states which are in the kernel of (i.e.~annihilated by) $\mathcal{Q}^\dagger$.
Indeed, it can be shown that the kernel of the effective supercharge $\mathcal{Q}^\dagger$ directly follows from
the kernel of the original supercharge $Q^\dagger$.
To see this, it suffices to repeat the Born-Oppenheimer procedure for an $r$-fermion state annihilated by $Q^\dagger$,
(namely, $Q_\alpha^\dagger|\psi^{(r)}\rangle=0$); so as to obtain
\begin{equation}
(\mathcal{Q_\alpha}^{(r)})^\dagger_{\sigma}\psi^{(r)}_\sigma=0,
\end{equation}
where $\psi^{(r)}_\sigma$ is the corresponding Born-Oppenheimer wavefunction, defined as
\begin{equation}
\psi^{(r)}_\sigma(M)\equiv\langle M, 0(M),\sigma|\psi^{(r)}\rangle,
\end{equation}
with $\sigma$ labeling the ground state degeneracy of the fermionic part of the wavefunction.
The examination of other multiplets is thus reduced to the study of the subset of wavefunctions in the kernel of $Q^\dagger$
that has a nonzero fermion number.
In accordance to equation (\ref{QQ}), the bosonic part $\phi$ of these wavefunctions must satisfy
\begin{equation}
\left(g^2\frac{\partial}{\partial M_{ia}}+\frac{1}{2}\epsilon_{ijk}F_{jk}^a\right)\phi (M)=0.
\label{kernel}
\end{equation}
It can be easily seen that such states are of the form
\begin{equation}
\phi(M)\sim \textrm{exp}\Big[-\frac{1}{g^2}\left(\frac{\text{Tr}(M^TM)}{2\rho}-\text{det}(M)\right)\Big].
\label{zeromode}
\end{equation}
Due to the unbounded term $\text{det}(M)$ in the above exponential, such wavefunctions are in general non-normalizable.
However, for a very small radius of $S^3$, the quadratic term  $\text{Tr}(M^TM)$ in the exponential dominates
and $\phi$ approaches a Gaussian with a sharp peak.
Therefore, in the limit $\rho<<1$, other supersymmetry multiplets arise in the spectrum. 

\section{Conclusion}
\label{sec5}

By pulling back the set of left-invariant connections of the full Yang-Mills theory onto the real superspace,
we obtain a natural quantum mechanical matrix model reduction of the $\mathcal{N}=1$ super-Yang-Mills gauge multiplet.
We then examine the spectrum of the corresponding Hamiltonian,
which is that of the matrix model for $SU(2)$ Yang-Mills theory coupled to adjoint fermions.
We proceed to quantize our model in the Born-Oppenheimer approximation,
by treating the gauge fields as slow degrees of freedom and the gauginos as fast ones.
This leads to two distinct phases for the matrix model: a color-spin-locked phase at the boundary and a bulk phase.

The apparently supersymmetry-violating quantization scheme we use recovers supersymmetry in an interesting and subtle way.
The spectra of the effective Hamiltonians in the different sectors of the Hilbert space
---corresponding to the fermions filling different numbers of Fermi energy levels---, get related by operators called effective supercharges. As a result, the different effective Hamiltonians organize themselves into multiplets, with the spectra related as (\ref{eq410}).
Supersymmetry is thus restored in the full Hilbert space, even though it is lost in any one sector.
Each and every sector is sensitive to the non-trivial quantum phase structure.
This can be most easily verified by noting that the effective supercharges exhibit a similar singular behavior
as the effective potential (\ref{phiasP}) for the gauge dynamics when one approaches the boundary from the bulk and vice-versa. 
We observe that there naturally exists two types of multiplets of effective Hamiltonians:
one starting with the purely bosonic (or fermion vacuum) sector and one starting with the hole vacuum sector.
The study of other multiplets leads to non-normalizable states, unless we work on a spatial sphere of very small radius. 

A simple yet interesting generalization of our matrix model consists on its coupling to a Wess-Zumino matter multiplet,
with the fermions and scalar field transforming in the fundamental representation of the gauge field.
As noted in~\cite{Pandey:2016hat}, fundamental fermions are sensitive to a wider variety of phases.
In particular, a special corner of the gauge configuration space arises as a separate phase of the theory,
which corresponds to rank one matrices $M$, such that $\text{det}(M)=0$.
Normalizable Gaussian solutions to (\ref{kernel}) exist only in this phase,
where we expect to obtain additional supersymmetry multiplets in the spectrum compared to the other phases.

\vspace{0.5cm}

{\bf Acknowledgements:}
We are grateful to N.~Acharyya for enlightening conversations in the early stages of the project.
We'd like to thank A.~P.~Balachandran, D.~O'Connor and M.~Hanada for useful discussions.
Angnis Schmidt-May was kind enough to read through an earlier version of this work.
The work of VED is supported by a grant from the Max Planck Society.

\end{document}